\documentclass[a4paper,12pt]{article}
\usepackage[utf8]{inputenc}
\usepackage{amssymb, mathtools}
\usepackage{algpseudocode,algorithm}
\usepackage[margin=2.00cm]{geometry}
\usepackage{tabularx}
\usepackage[numbers]{natbib}
\usepackage{setspace}
\usepackage{fancybox}
\usepackage{tikz}
\usepackage{rotating}
\usepackage{makecell}

\newcommand{\wrt}[1]{\mathrm{d}{#1}}
\newcommand{\bs}[1]{\boldsymbol{#1}}

\title{Efficiently simulating discrete-state models with binary decision trees}
\author{Christopher Lester\footnote{Mathematical Institute, Woodstock Road, Oxford, OX2 6GG. Email: \texttt{lesterc@maths.ox.ac.uk}.}~, Ruth E. Baker\footnote{Mathematical Institute, Woodstock Road, Oxford, OX2 6GG.}~, Christian A. Yates\footnote{Department of Mathematical Sciences, North Rd, Bath BA2 7AY.}~}
\date{16 January 2020}

\begin{document}

\maketitle

 \begin{abstract}
 
 Stochastic simulation algorithms (SSAs) are widely used to numerically investigate the properties of stochastic, discrete-state models. The Gillespie Direct Method is the pre-eminent SSA, and is widely used to generate sample paths of so-called agent-based or individual-based models. However, the simplicity of the Gillespie Direct Method often renders it impractical where large-scale models are to be analysed in detail. In this work, we carefully modify the Gillespie Direct Method so that it uses a customised binary decision tree to trace out sample paths of the model of interest. We show that a decision tree can be constructed to exploit the specific features of the chosen model. Specifically, the events that underpin the model are placed in carefully-chosen leaves of the decision tree in order to minimise the work required to keep the tree up-to-date. The computational efficiencies that we realise can provide the apparatus necessary for the investigation of large-scale, discrete-state models that would otherwise be intractable. Two case studies are presented to demonstrate the efficiency of the method. 
  
 \end{abstract}

\section{Introduction} \label{__label__2f26e5d90d5f4b5e8e26a21643746bad}
In the life sciences, models are extensively used to understand and interrogate data, thereby playing an important part in furthering our understanding of a range of biological mechanisms~\citep{__ref__33611e8f2913412d9e8fe5864f57d3c7, __ref__1a3541a1fee44aae857a8870b87649ef}. A multitude of stochastic modelling approaches have enjoyed widespread successes as interpretable models of real-world physical phenomena~\citep{__ref__221389c689d04b34b5c22896abc672e7}. In this manuscript, we focus on \emph{discrete-state} models that explicitly track each event that changes the model's state. Discrete-state models take on many forms: by way of example, these include the biochemical reaction networks described by a Chemical Master Equation~\citep{__ref__33611e8f2913412d9e8fe5864f57d3c7, __ref__1a3541a1fee44aae857a8870b87649ef, __ref__221389c689d04b34b5c22896abc672e7}, and the agent- or individual-based models designed to undertake detailed epidemiological investigations~\citep{__ref__6b94d95a4d484ce0b244336a44f6ecea, __ref__f498db3c1e964163a8a1231349bce15e}. 

The dynamics of realistic, discrete-state models are often analytically intractable: specifically, it is typically impossible to obtain closed-form equations that probabilistically report even summary statistics of the system of interest~\citep{__ref__1a3541a1fee44aae857a8870b87649ef, __ref__221389c689d04b34b5c22896abc672e7}. To make sense of the dynamics of a chosen model, it is often necessary to use Monte Carlo simulation to generate an ensemble of sample paths: the sample paths are realisations, or example time-series, of the model. Once the sample paths have been generated, summary statistics can be estimated to characterise the dynamics~\citep{__ref__115d94a32a21424e98958e07bcc499e4}. 

The Gillespie Direct Method (`GDM') is an event-driven algorithm that is widely used to generate sample paths of a chosen stochastic model. The GDM generates sample paths by sequentially sampling each event that takes place over a given time interval~\citep{__ref__115d94a32a21424e98958e07bcc499e4, __ref__95f69d794ca6465396aa1b23c5e7ad6b, __ref__35755be6131a4c9e83117b17737c061c}. Many variations of the GDM have been implemented, and a range of software packages can be employed to produce sample paths~\citep{__ref__db66f64ab877460bb7754f2fcb5a3f3c, __ref__2be02b1393a8492faf235127e91447fd}. To ensure that the largest, most complex and physically-relevant models can be thoroughly investigated -- that is, ensuring that parameter sweeps, model validation and verification studies can be carried out in sufficient detail -- the GDM must be studied, refined and improved upon. 

In this manuscript, we focus on one particular computational challenge encountered when using the GDM. Specifically, if there are $M$ classes of event that characterise the interactions (often described as reactions) within the model, then the standard GDM uses $\mathcal{O}(M)$ computational steps to select each event that takes place~\citep{__ref__115d94a32a21424e98958e07bcc499e4}. Each event modifies the populations of the model: therefore, the reaction or interaction rates need to be updated, and this requires a further $\mathcal{O}(M)$ computational steps. As explained in the previous paragraph, this computational complexity can preclude the in-depth analysis of large-scale models.

The GDM can be modified so that a binary decision tree is used to simulate each event~\citep{__ref__c70bde5c8fa14266b1df999c5ac5bd04}. With a decision tree, an average of $\mathcal{O}(\log M)$ steps are required to generate each reaction or interaction, but if the decision tree is not carefully chosen, a significant number of computational steps will still be required as a consequence of the reaction or interaction rates needing to be updated.

As there is a remarkable degree of flexibility in the choice of the binary decision tree that is used, we take the opportunity to design a bespoke binary decision tree that can be quickly updated after each interaction is simulated. We show that, with a judicial decision tree choice, sample paths can be generated with a substantially lower level of computational resources.

\subsection{Outline} \label{__label__febf33ffe4ef4976ab592234a6f77b15}
This manuscript is arranged as follows: in Section \ref{__label__61c47b66345c4b9799cc74c652be3334}, we provide background material so that a discrete-state model is carefully defined. In Section \ref{__label__28abc88440d1434db601b26de0e65d7f}, we develop a bespoke decision tree-based simulation algorithm. Our new algorithm is carefully tested, and the results are presented in Section \ref{__label__d8606a0a95494d729a9d596e0819d59d}. A discussion of the findings is presented in Section \ref{__label__47c2e02981814fe88813dc92598acc9a}.

\section{Background information}  \label{__label__61c47b66345c4b9799cc74c652be3334}
This section provides the background material that underpins the rest of the manuscript. We start by introducing notation, and then proceed to outline the Gillespie Direct Method and its derivatives. 

\subsection{Notation and conventions} \label{__label__a2fc002242ec43d68d5758981b69ae1b}
A discrete-state model comprises individuals, each of a given `species'. The population of each species can change due to interactions that are often referred to as  `reactions'. We provide a description of the terms `species' and `reactions' as follows:

\textbf{Species.} We use a vector $\boldsymbol{X}(t)$ to denote the `state vector' that records the populations comprising the model. There is considerable flexibility in how the vector $\boldsymbol{X}(t)$ might be defined:

\begin{itemize}
 \item if a model is spatially homogeneous, then the \emph{total population} of each species can be tracked through time;
 \item if a model is spatially inhomogeneous, then the location of different individuals must be recorded. One approach involves discretising the region of interest into $K$ `patches' or  `compartments'~\citep{__ref__0006fcbb9b854a86a6496bf60d524597}, which we enumerate as $1, 2, \dots, K$. In this case, the \emph{patch population} of each species in each patch is recorded. 
\end{itemize}

\textbf{Reactions.} Any interaction, or change to the state vector, is as a result of a `reaction'. We index reactions as $1, 2, \dots, M$. The propensity of a reaction specifies its average rate; it is common for the propensity to be calculated according to mass action kinetics~\citep{__ref__1a3541a1fee44aae857a8870b87649ef}. In addition, each reaction is associated with a stoichiometric vector that indicates the net change to the state vector, $\boldsymbol{X}$, after the reaction has `fired'.

We denote the propensity function of the $j$-th reaction as $a_j(\boldsymbol{X})$, and the stoichiometric vector as $\nu_j$. Thus, if the $j$-th reaction takes place, we update $\boldsymbol{X}$ by setting \begin{equation*}\boldsymbol{X} \leftarrow \boldsymbol{X} + \nu_j.\end{equation*}

\textbf{Example.} For an S-I-S compartment model~\citep{__ref__f498db3c1e964163a8a1231349bce15e}, we use `$S_k$' to denote a susceptible individual in patch $k$; for an infected individual in patch $k$, we write `$I_k$'. For a $K$-patch model the state vector, $\boldsymbol{X}$, can be expressed as \begin{equation*}
\boldsymbol{X} = [S_1, I_1, S_2, I_2, \dots, S_K, I_K]^T.
\end{equation*} 

Infection takes place within a patch: infection of a susceptible individual in patch $k$ by contact with an infected individual (also within patch $k$) can be represented by the reaction $$S_k + I_k \rightarrow 2 \cdot I_k,$$ whilst recovery of an infected individual can be represented as $$I_k \rightarrow S_k.$$ The former reaction reduces the $S_k$ population by one, and increases the $I_k$ population by one. The latter reaction has the reverse effect. For the former reaction, the propensity is proportional to $S_k \cdot I_k$, whilst for the latter reaction, the propensity is proportional to $I_k$. The constants of proportionality are known as rate constants and may be estimated from experimental data~\citep{__ref__1a3541a1fee44aae857a8870b87649ef}. 

If individuals are able to move from one patch to another these movements are treated as reactions and can be represented as $$I_k \rightarrow I_j,\,\,\,\,\, S_k \rightarrow S_j,$$ for distinct patches $k$ and $j$. For the former reaction, the $I_k$ population reduces by one, and the $I_j$ population increases by one. A similar result applies for the latter reaction. The propensity of the former reaction is proportional to $I_k$, whilst for the latter reaction, the propensity is proportional to $S_k$. The constants of proportionality can be calculated by referring to the diffusion coefficients.\hfill $\blacksquare$

\subsection{Gillespie Direct Method} \label{__label__86b1e5bb048a42bbbea56dafad702ac4}
The GDM is a serial method in the sense that it generates sample paths of a model of interest by simulating individual reactions in the order in which they take place. A key benefit of this approach is that it can be implemented in a straightforward manner, as shown in Algorithm \ref{__label__b05935eb781d4c5a83ff2f5c80d584bb}. 

\begin{algorithm}[bth]
\caption{Gillespie Direct Method.\protect\vphantom{$A_A^A$}}
\label{__label__b05935eb781d4c5a83ff2f5c80d584bb}
 \begin{algorithmic}[1]
  \Require initial conditions, $\boldsymbol{X}(0)$, and final time, $T$.\protect\vphantom{$A_A^A$}
  \State set $t \leftarrow 0$
  
  \Loop
  \State for each reaction $j = 1, \dots, M$, calculate $a_j(\boldsymbol{X})$ and set $a_0(\boldsymbol{X}) \leftarrow \sum_{j=1}^M a_j (\bs{X})$
  \State sample $\Delta \leftarrow \text{Exp}(a_0)$  
  \If{$t + \Delta > T$}
  \State \textbf{break}
  \EndIf
  \State choose the $m$-th reaction to fire: each reaction $j$ is chosen with probability $a_j / a_0$
  \State set $\boldsymbol{X} \leftarrow \boldsymbol{X} + \nu_m$, and set $t \leftarrow t + \Delta$
  \EndLoop
 \end{algorithmic}
\end{algorithm}

At each iteration of the loop of Algorithm \ref{__label__b05935eb781d4c5a83ff2f5c80d584bb}, the propensity (or average rate) of each reaction is calculated, and the propensities are summed together to give the total propensity (i.e. the average rate at which any reaction fires). Then: \begin{itemize}
                                                                                                                                                                                                                                                  \item in line 4, an exponential variate is used simulate the time between successive reactions;
                                                                                                                                                                                                                                                  \item in line 8, a specific reaction is chosen. The inverse transform method is a popular technique for choosing a specific reaction, and this method proceeds as follows. First, sample $u$ from a uniform random variate on $(0,1)$. Then determine the index $k$, where $k$ satisfies \begin{equation*} 
\sum_{j = 1}^{k-1} a_j (\boldsymbol{X}) \le u \cdot a_0(\boldsymbol{X}) < \sum_{j = 1}^{k} a_j(\boldsymbol{X}).
\end{equation*} In practice, one can follow a `bottom-up' search procedure: set $k \leftarrow 1$, and then increase $k$ until $\sum_{j=1}^k a_j(\boldsymbol{X}) > u \cdot a_0(\boldsymbol{X})$. If the reactions are randomly ordered, then this search procedure requires an average of $M/2$ steps to complete. 

\end{itemize}

\subsection{Optimising the Gillespie Direct Method} \label{__label__3b9998ca3c6443c88ec69bc047871c44}
We now present simulation methods that are variants of the GDM. Each variant that we highlight addresses specific deficiencies of Algorithm \ref{__label__b05935eb781d4c5a83ff2f5c80d584bb}. We will use these improved simulation methods as inspiration for the new framework outlined in Section \ref{__label__28abc88440d1434db601b26de0e65d7f}. 

\textbf{Optimised and Sorting Direct Methods.} \citet{__ref__a8995c057f7c4f3cb23c698d718f24ac} describe the Optimised Direct Method. When compared with the GDM, the method is more computationally efficient for two principal reasons. Firstly, at each iteration of the GDM, all propensity values are recalculated, even though some of these values will remain the same. Instead \citet{__ref__a8995c057f7c4f3cb23c698d718f24ac} draw up a dependency graph to indicate which propensity values might need to be updated at each iteration of the algorithm. Those propensities that are unaffected by the chosen reaction are not recalculated. Secondly, the Optimised Direct Method uses a carefully-chosen `bottom-up' inverse transform method used to select reactions. An average of $\sum_{j=1}^M j \cdot a_j(\boldsymbol{X}) / a_0(\boldsymbol{X})$ steps are required to generate each value of $k$. If the propensities are sorted so that $a_1 \ge a_2 \ge \dots \ge a_M$ then the average number of steps required to generate $k$ is minimised.

\citet{__ref__e9a0ea25095f41ef97d1cdf874d0eaeb} present an algorithm known as the Sorting Direct Method. This method takes into account the fact that reaction frequencies can change over the time-span of interest. \citet{__ref__e9a0ea25095f41ef97d1cdf874d0eaeb} dynamically reorder the reactions in a specified list that is used as part of the inverse transform method to select reactions. Every time a reaction is fired, it swaps places with the reaction above it in the list. The effect is that frequently-encountered reactions appear towards the top of the list, whilst rarely-seen reactions linger at the bottom. 

\textbf{Logarithmic Direct Method.} If the frequencies of different reactions vary by orders of magnitude, then the above-mentioned methods can significantly accelerate stochastic simulation. However, if all reactions occur with similar frequencies, then the methods provide only limited benefits. Moreover, the computational complexity of the method for choosing the reaction to fire remains $\mathcal{O}(M)$. In a technical report, \citet{__ref__c70bde5c8fa14266b1df999c5ac5bd04} point out that propensity values are repeatedly summed -- first, to calculate the total propensity, $a_0(\boldsymbol{X})$, and second, as part of the inverse transform method. \citet{__ref__c70bde5c8fa14266b1df999c5ac5bd04} therefore propose the following strategy: instead of storing the propensities, store the cumulative  sum of the first $j$ (for $j = 1, \dots, M$) propensities, $b_1, \dots, b_M$, where $b_j \equiv \sum_{j' = 1}^j a_j(\boldsymbol{X})$ (with $b_0 \equiv 0$; equivalently, set $b_j \equiv b_{j-1} + a_j(\boldsymbol{X})$). Then, a reaction $k$ is fired by performing a binary search to determine the value of $k$ such that  \begin{equation*} 
b_{k-1} < u \cdot b_M < b_k.
\end{equation*} The computational complexity of the search depth of this method is $\mathcal{O}(\log{(M)})$; however, $\mathcal{O}(M)$ steps are still required to generate $b_1, \dots, b_M$.

\textbf{Recycling random numbers.} The unmodified GDM uses two random numbers during each iteration: one random number to determine the time spanned by that step, and a second random number as part of the inverse transform method to choose a reaction. \citet{__ref__66d52925a8c64bbc83068374f2ed9600} noted that sample paths can be generated by using only one random number at each step. They show that if a $(0,1)$-uniform random variate, $u$, is sampled and the inverse transform method is used to sample a reaction $k$, then $v$, defined as
\begin{equation*}
v \equiv \begin{cases}
\dfrac{u\cdot a_0(\boldsymbol{X})}{a_1(\boldsymbol{X})}, & k = 1, \\
\dfrac{u \cdot a_0(\boldsymbol{X})-\sum_{i=1}^{k-1} a_i(\boldsymbol{X})}{a_{k}(\boldsymbol{X}) }, & k > 1,
\end{cases}
\end{equation*} is a $(0,1)$-uniform random variate that is independent of $k$. The random variate $v$ can be used to generate the waiting time until the next reaction, $\Delta$, as
\begin{equation*}
 \Delta \equiv  \dfrac{\log(1 / v)}{a_0(\boldsymbol{X}) }.
\end{equation*}

We are now in a position to set out a general approach to efficiently simulating a large-scale stochastic model in more detail. 

\section{Methods and developments}  \label{__label__28abc88440d1434db601b26de0e65d7f}
In this section, we present a new, integrative method for generating sample paths of the models specified in Section \ref{__label__a2fc002242ec43d68d5758981b69ae1b}. Building on the GDM, we efficiently simulate reactions with a binary decision tree. We demonstrate that a well-chosen decision tree can lead to substantial computational savings, and we provide a heuristic method for choosing such decision trees.

\subsection{Binary decision tree} \label{__label__c29961ca9a504cb88e8946d21fe7a916}
\begin{figure}[ht]

\hrulefill
\vspace{1mm}
 \centering
 \centering

 \begin{minipage}[c]{.49\textwidth}\centering
   
 \begin{tikzpicture}[sibling distance=30.42mm, edge from parent/.style={draw,-latex}]
 \node { \textbf{\color{blue}root}}
     child { node {\textbf{\color{blue}node}} 
     child { node {\textbf{\color{olive}leaf}}} 
     child { node {\textbf{\color{olive}leaf}}}
     }
     child { node {\textbf{\color{olive}leaf}}
     };  
 \end{tikzpicture}
\end{minipage}\hfill\begin{minipage}[c]{.49\textwidth}
 \begin{tikzpicture}[sibling distance=30.42mm, edge from parent/.style={draw,-latex}]
 \node {$\omega_5 = \omega_3 + \omega_4$}
     child { node {$\omega_4 = \omega_1 + \omega_2$} 
     child { node {$\omega_1 $}} 
     child { node {$\omega_2 $}}
     }
     child { node {$\omega_3 $}
     };  
 \end{tikzpicture}
\
 \end{minipage}

\caption{An example of a binary tree. On the left-hand side, the `leaf' nodes are shown in an olive colour, whilst the `non-leaf' nodes are shown in blue. The weights of the nodes are shown on the right-hand side .} \label{__label__3d5bf840aa14424d911f3f5d39eaca07}

\hrulefill

\end{figure}
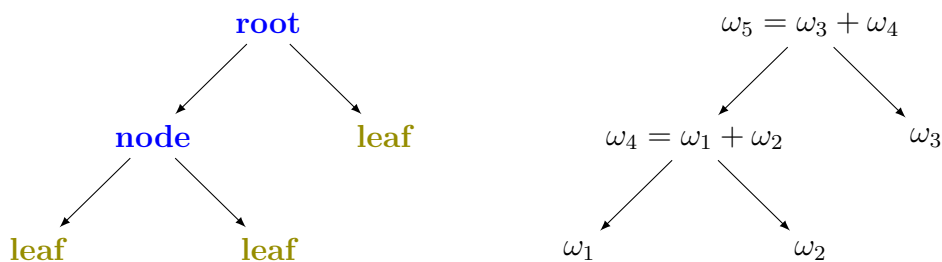

The implementation of the GDM described in Section \ref{__label__86b1e5bb048a42bbbea56dafad702ac4} employed the inverse-transform method to fire (i.e. sample) reactions. Taking into account the Logarithmic Direct Method outlined in Section \ref{__label__3b9998ca3c6443c88ec69bc047871c44}, we now detail how a binary decision tree can be used to fire reactions efficiently. We start with a definition:

\textbf{Binary tree.} Whilst more formal definitions are possible, a binary tree, $D$, comprises a set of nodes that are joined together by edges, subject to the following: \begin{itemize}
\item a node is classified as a `leaf' or `non-leaf';
\item a `non-leaf' is the `parent-node' to precisely two `child-nodes': a left-child, and a right-child;
\item a `leaf' node has no children;
\item there is exactly one node which has no `parent-node': this is the `root' of the tree.
\end{itemize} An edge connects two nodes if there is a parent-child relationship between them; an example of a binary tree is shown in Figure \ref{__label__3d5bf840aa14424d911f3f5d39eaca07}. We will assign a weight $\omega$ to each node $\alpha$. The descendants of a node include the child-nodes, the child-nodes of the child-nodes, and so on.

\textbf{Binary trees as a decision tool.} We use a binary tree to simulate reactions as follows. We represent each reaction, $j = 1, \dots, M$, as a leaf node in the binary tree $D$, with weight $\omega = a_j(\boldsymbol{X})$. We then generate non-leaf nodes to join the tree up. The weight of each non-leaf node is given by the sum of the weights of its child-nodes. In particular, this means the weight of a node is equal to the total propensity of all of the leaves that are its descendants. The weight of the root node is therefore equal to the sum of all the propensities (i.e. the total propensity, $a_0(\boldsymbol{X})$).

This means that a reaction can be chosen by following a recursive procedure: start from the root, and select the left-child with probability $\omega_{\text{left}} / (\omega_{\text{left}} + \omega_{\text{right}})$ (where $\omega_{\text{left}}$ and $\omega_{\text{right}}$ refer to the respective weights of the child-nodes). Otherwise, select the right node. We are thereby comparing the total propensity associated with the descendants of the left-child with that of the right-child so as to decide which branch of the tree to explore further. Now consider the children of the node just selected (i.e. the `grand-children' of the root node): select the left (grand-) child with probability $\omega'_{\text{left}} / (\omega'_{\text{left}} + \omega'_{\text{right}})$, otherwise select the right (grand-) child. Recursively repeat until a leaf is selected. The reaction associated with this leaf is fired.

We illustrate further with an example.

\textbf{Example.} Consider an S-E-I-R compartment model~\citep{__ref__f498db3c1e964163a8a1231349bce15e}.  In this model, a susceptible individual (state `S'), is exposed to a disease (state `E') after coming into contact with an infectious (state `I') individual. After a random period of time, exposed individuals become infectious, and infectious individuals are removed (state `R') from the system (either through recovery or death). 

\begin{figure}[ht]

\hrulefill
\vspace{1mm}
 \centering

 \begin{minipage}[c]{.49\textwidth}\centering
 \begin{align*}
    \alpha =  1: \hspace{12pt} & S + I \xrightarrow{\theta_1} E + I \\
    \alpha =  2: \hspace{12pt} & E \xrightarrow{\theta_2} I \\
    \alpha =  3: \hspace{12pt} & I \xrightarrow{\theta_3} R  \\
   \end{align*}

\end{minipage}\hfill\begin{minipage}[c]{.49\textwidth} \centering
 \begin{tikzpicture}[sibling distance=30.42mm, level distance=21.0mm, edge from parent/.style={draw,-latex}]
 \node [rectangle,draw] {\makecell[c]{$\alpha = 5$\\$(\omega_5 = \omega_3 + \omega_4)$}}
     child { node [rectangle,draw]{\makecell[c]{$\alpha = 4$\\$(\omega_4 = \omega_1 + \omega_2)$}} 
     child { node [rectangle,draw]{\makecell[c]{ $\alpha = 1$\\$(\omega_1 = \theta_1 \cdot S \cdot I)$}}} 
     child { node [rectangle,draw]{\makecell[c]{ $\alpha = 2$\\$(\omega_2 = \theta_2 \cdot E)$}}}
     }
     child { node [rectangle,draw]{\makecell[c]{ $\alpha = 3$\\$(\omega_3 = \theta_3 \cdot I)$}}
     };  
 \end{tikzpicture}

\end{minipage}
\caption{The reactions of an S-E-I-R model are shown on the left, and a corresponding binary decision tree is shown on the right. The index of each node is shown, together with its weight in brackets. Further details are contained within the main text.} \label{__label__d1785f09648847228c024cec53491ebd}

\hrulefill

\end{figure}
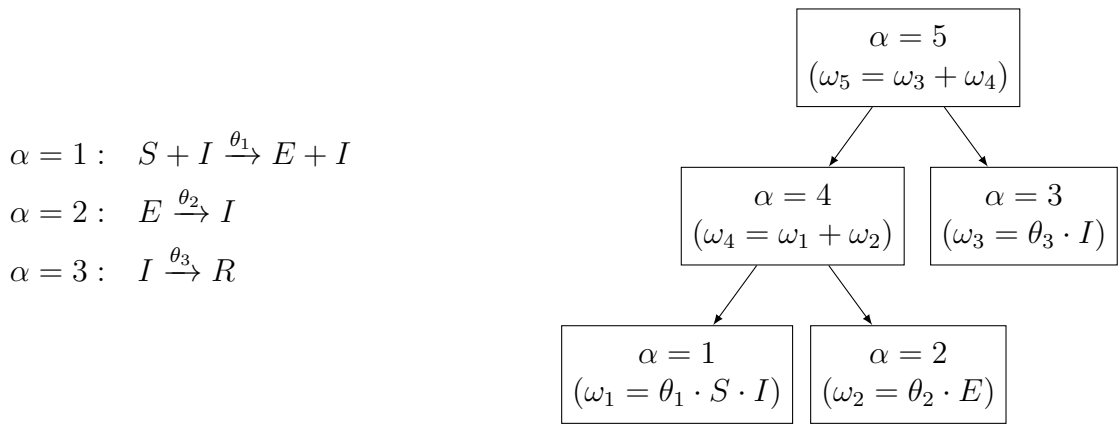

The reactions that govern this model are detailed on the left of Figure \ref{__label__d1785f09648847228c024cec53491ebd}. On the right, a binary decision tree is shown. The reactions are represented by leaves $\alpha = 1, 2, 3$. The other nodes (created to connect the tree) are represented by $\alpha = 4, 5$. The weights of $\alpha = 1, 2$ and $3$, $\omega_1, \omega_2$ and $\omega_3$, are given by the propensities of their respective reactions: $\theta_1\cdot S \cdot I$, $\theta_2\cdot E$, and $\theta_3\cdot I$, respectively. The weight of $\alpha = 4$ is $\omega_4 = \omega_1 + \omega_2$, and the weight of $\alpha = 5$ is $\omega_5 = \omega_4 + \omega_3$. $\alpha = 5$ is the root.

Having described how one might modify the GDM to make use of a binary decision tree, we now use Section \ref{__label__a26ebd50833f4a31a2b48911100b9736} to provide pseudo-code of our simulation algorithm. In Section \ref{__label__fcf2cdd712cb4f66aac6cf276af7827b}, we explain how the tree $D$ can be chosen to ensure computational efficiency. 

\subsection{A reformulated Gillespie Direct Method} \label{__label__a26ebd50833f4a31a2b48911100b9736}

A redesigned GDM is presented as Algorithm \ref{__label__04d1a41b281d4dea82aaf962876f1a77}. Algorithm \ref{__label__04d1a41b281d4dea82aaf962876f1a77} will use a binary decision tree to fire reactions. In turn, Algorithm \ref{__label__04d1a41b281d4dea82aaf962876f1a77} relies on three sub-routines: Algorithm \ref{__label__21cff712335d4bf78185004ef71eacdb} is used for the initial parametrisation of the decision tree, Algorithm \ref{__label__e371a49d18db455eaac0aef43c64a9f9} is directly responsible for firing reactions, and Algorithm \ref{__label__a13dc06ab0a6434289b4a43f9f6e09cd} ensures the tree remains up-to-date. Note that Algorithm \ref{__label__21cff712335d4bf78185004ef71eacdb} runs once: thereafter, we alternate between Algorithms \ref{__label__e371a49d18db455eaac0aef43c64a9f9} and \ref{__label__a13dc06ab0a6434289b4a43f9f6e09cd}. We first discuss Algorithms \ref{__label__e371a49d18db455eaac0aef43c64a9f9} and \ref{__label__a13dc06ab0a6434289b4a43f9f6e09cd}:

\textbf{Firing reactions.} In Section \ref{__label__c29961ca9a504cb88e8946d21fe7a916} we explained that, by starting at the root of the tree $D$, and repeatedly choosing either the left or right child-node, eventually a leaf is reached. The reaction associated with this leaf is fired. At each step of Algorithm \ref{__label__e371a49d18db455eaac0aef43c64a9f9}, the total propensity associated with all reactions descending the left- and the right-child-nodes are compared. The left-child is selected with probability $\omega_{\text{left}} / (\omega_{\text{left}} + \omega_{\text{right}})$. A $(0,1)$-uniform random variate, $u$, is generated. Then,  \begin{equation*}
u \le \dfrac{\omega_{\text{left}}}{\omega_{\text{left}} + \omega_{\text{right}}} \implies \text{ go left}, \hspace{.3in} u > \dfrac{\omega_{\text{left}}}{\omega_{\text{left}} + \omega_{\text{right}}} \implies \text{ go right}.
\end{equation*} This procedure can be repeated, using a new random number at each step, until a leaf (i.e. a reaction) is reached.  

The algorithm is more efficient if the random numbers are recycled. To do this, let the random number used at step $i$ be $u_{i}$. Then we let $u_{i+1}$, the random number to be used at the next step, be chosen as \begin{equation*}
\text{picked left} \implies u_{i+1} \equiv  \dfrac{ u_i \cdot (\omega_{\text{left}} + \omega_{\text{right}})}{\omega_{\text{left}}}, \hspace{.3in} \text{picked right}  \implies u_{i+1} \equiv \dfrac{u_i\cdot(\omega_{\text{left}} + \omega_{\text{right}}) - \omega_{\text{left}}}{\omega_{\text{right}}}.
\end{equation*} Clearly, $u_{i}$ and $u_{i+1}$ depend on each other. However, if we left $d_i \in \{\ell, r\}$ indicate the choice of picking the left- or right-child at step $i$, then the decision $d_{i+1}$ is independent of decision $d_i$: \begin{align*}
\mathbb{P}\Big[d_{i+1}, d_{i}\Big] & = \int \mathbb{P}\Big[d_{i+1}, d_{i} \mid u_{i+1}\Big] \hphantom{\cdot} \wrt{\mathbb{P}\Big[ u_{i+1}\Big]}  \\
& = \int \mathbb{P}\Big[d_{i+1} \mid d_i, u_{i+1} \Big] \cdot \mathbb{P}\Big[d_{i} \mid u_{i+1}\Big] \hphantom{\cdot} \wrt{\mathbb{P}\Big[u_{i+1}\Big]} \\  
& = \int \mathbb{P}\Big[d_{i+1} \mid u_{i+1} \Big] \cdot \mathbb{P}\Big[d_{i} \Big] \hphantom{\cdot} \wrt{\mathbb{P}\Big[u_{i+1}\Big]} \\
& = \mathbb{P}\Big[d_{i+1}\Big] \cdot \mathbb{P}\Big[d_{i}\Big].
\end{align*} If, instead of working with $(0,1)$-uniform random variates we use $v \sim (0, \omega_{\text{left}} + \omega_{\text{right}})$-uniform random variates, then we explore $D$ as follows: \begin{equation*}
v \le {\omega_{\text{left}}}\implies \text{ go left}, \hspace{.3in} v > \omega_{\text{left}} \implies \text{ go right}.
\end{equation*} Random numbers can be recycled, and we can take $v$ as  \begin{equation*}
\text{picked left} \implies v_{i+1} \equiv v_i, \hspace{.3in} \text{picked right}  \implies v_{i+1} \equiv {v_i - \omega_{\text{left}}}.
\end{equation*} Algorithm \ref{__label__e371a49d18db455eaac0aef43c64a9f9} implements this method.

\textbf{Updating the tree.} After a reaction $j$ has fired, the population numbers change, and it could be necessary for some reaction propensities to be updated. For each reaction, $j$, a list of propensities that need to be recalculated can be drawn up \emph{a priori} as\begin{equation*}
K_j = \big\{ k : \text{ at least one of the reactants of reaction $k$ is changed by reaction $j$} \big\}.
\end{equation*} After a reaction $j$ has fired, all the leaf nodes of $D$ that are included in the list $K_j$ need their weights updated. In addition, as the weight of each non-leaf node is given by the sum of the weights of its child-nodes, a number of other nodes will need their weights recalculated. We write $\varLambda_j$ as the list of all nodes that need their weights recalculated. Algorithm \ref{__label__a13dc06ab0a6434289b4a43f9f6e09cd} updates the tree. Figure \ref{__label__f2d9dfa6650446a4ad861136457bf8a6} (which is contained within the next section) provides a diagrammatical representation of this update procedure.

\begin{algorithm}[p]
\caption{Efficient Gillespie Direct Method.\protect\vphantom{$A_A^A$}}
\label{__label__04d1a41b281d4dea82aaf962876f1a77}

 \begin{algorithmic}[1]
  \Require initial conditions, $\boldsymbol{X}(0)$, decision tree, $D$, and final time, $T$.\protect\vphantom{$A_A^A$}
  \State set $t \leftarrow 0$
  \State for each reaction $j = 1, \dots M$, calculate propensity $a_j(\boldsymbol{X})$
  \State populate tree $D$: use \textsc{Algorithm } \ref{__label__21cff712335d4bf78185004ef71eacdb}
  \Loop
  \State sample $\Delta \leftarrow \text{Exp}(\omega_{\text{root}})$
  \If{$t + \Delta > T$}
  \State \textbf{break}
  \EndIf
  \State choose the $k$-th reaction to fire: use \textsc{Algorithm } \ref{__label__e371a49d18db455eaac0aef43c64a9f9}
  \State set $\boldsymbol{X} \leftarrow \boldsymbol{X} + \nu_k$, and set $t \leftarrow t + \Delta$
  \State update tree $D$ as per \textsc{Algorithm } \ref{__label__a13dc06ab0a6434289b4a43f9f6e09cd}.
  \EndLoop

 \end{algorithmic}
\end{algorithm}

\renewcommand{\thealgorithm}{2a}
\begin{algorithm}[p]
\caption{Sub-method to populate decision tree with weights.\protect\vphantom{$A_A^A$}}
\label{__label__21cff712335d4bf78185004ef71eacdb}

 \begin{algorithmic}[1]
 \Require Tree $D$ \protect\vphantom{$A_A^A$}
  \For{each node $\alpha \in D$}, ordered by decreasing distance from root \protect\vphantom{$A_A^A$}
  \If {$\alpha$ is a leaf}, set weight $\omega_\alpha \leftarrow a_j(\boldsymbol{X})$
  \Else { set weight $\omega_\alpha \leftarrow \omega_{\text{left}} + \omega_{\text{right}}$}.
  \EndIf
  \EndFor

 \end{algorithmic}
\end{algorithm}

\renewcommand{\thealgorithm}{2b}
\begin{algorithm}[p]
\caption{Sub-method to search for a reaction to fire. This algorithm is recursive, so a procedure is used to start.\protect\vphantom{$A_A^A$}}
\label{__label__e371a49d18db455eaac0aef43c64a9f9}

 \begin{algorithmic}[1]
\Require Binary Tree, $D$, and $r \sim U(0, \omega_{\text{root}})$. \protect\vphantom{$A_A^A$} 
\Procedure {Select Reaction}{}
\State \Return {\textsc{Decide}($\alpha_{\text{root}}$, $r$)}
\EndProcedure

 \end{algorithmic}
 \hrulefill
 
 \begin{algorithmic}[1]
\Require node, $\alpha$, with weight $\omega_\alpha$; left-child-node $\alpha_{\text{left}}$ with weight $\omega_{\text{left}}$; right-child-node $\alpha_{\text{right}}$\protect\vphantom{$A_A^A$}
\Function{Decide}{$\alpha, r$}:
\If{$\alpha$ is a leaf} \Return{$\alpha$} 
\Else
\If{$w_{\text{left}} > r$} \Return{\textsc{Decide}($\alpha_{\text{left}}$, $r$)}
\Else \hspace{1pt}
\Return{ \textsc{Decide}($\alpha_{\text{right}}$, $u - \omega_\text{left}$)}
\EndIf
\EndIf
\EndFunction

 \end{algorithmic}
\end{algorithm}

\renewcommand{\thealgorithm}{2c}
\begin{algorithm}[p]
\caption{Sub-method that updates decision tree with new weights, as required.\protect\vphantom{$A_A^A$}}
\label{__label__a13dc06ab0a6434289b4a43f9f6e09cd}

 \begin{algorithmic}[1]
 \Require Tree $D$ \protect\vphantom{$A_A^A$}, and reaction $j$
  \For{each node $\alpha \in \varLambda_j$}, ordered by decreasing distance from root \protect\vphantom{$A_A^A$}
  \If {$\alpha$ is a leaf}, set  $\omega_\alpha \leftarrow a_j(\boldsymbol{X})$
  \Else { set $\omega_\alpha \leftarrow \omega_{\text{left}} + \omega_{\text{right}}$}.
  \EndIf
  \EndFor

 \end{algorithmic}
\end{algorithm}

\subsection{Constructing binary trees} \label{__label__fcf2cdd712cb4f66aac6cf276af7827b}
We have thus established how a binary decision tree can be used to sample the reactions of a discrete-state stochastic model. We now consider what a highly efficient binary decision tree might look like. Once such a decision tree is decided upon, Algorithm \ref{__label__04d1a41b281d4dea82aaf962876f1a77} (which, in turn, calls Algorithms \ref{__label__e371a49d18db455eaac0aef43c64a9f9} and \ref{__label__a13dc06ab0a6434289b4a43f9f6e09cd}) will be utilised to generate sample paths of the model.

In Section \ref{__label__a26ebd50833f4a31a2b48911100b9736} we outlined the two major components of the computational cost of using a modified GDM: firstly, Algorithm \ref{__label__e371a49d18db455eaac0aef43c64a9f9} is used to sample each reaction by exploring the decision tree, and, secondly, Algorithm \ref{__label__a13dc06ab0a6434289b4a43f9f6e09cd} updates the decision tree as required. It is possible to directly optimise the performance of Algorithm \ref{__label__e371a49d18db455eaac0aef43c64a9f9} by using the Huffman coding algorithm. The Huffman algorithm is widely used in the field of communication theory, and places nodes with high firing probabilities close to the root, whilst placing nodes that fire rarely far from the root~\citep{__ref__2118077d287a48c89870a834c2690e86}. The effect of the Huffman algorithm is to minimise the number of steps required to sample a reaction, but it neglects the cost of updating the decision tree.

For a large class of event-based stochastic modelling problems, it is important to prioritise the task of minimising the cost of updating the decision tree. When a reaction is sampled, we traverse a \emph{single} path from the root of the tree to a node. When the tree is updated, we traverse \emph{multiple} paths from the leaves associated with reactions with new propensity values, up to the root. Following the notation of Section \ref{__label__a26ebd50833f4a31a2b48911100b9736}, if reaction $j$ fires, then $\|K_j\|$ propensities need to be recalculated, which means, potentially, $\|K_j\|$ paths to follow. Whilst each model has its own structure, for a model comprising many connected, interacting individuals, $\|K_j\|$ might typically be larger than one, so that the cost of updating the decision tree could dominate the cost of searching the tree. 

We seek to design a decision tree so that steps in the tree update process specified by Algorithm \ref{__label__a13dc06ab0a6434289b4a43f9f6e09cd} can be shared. For example, if the $j$-th reaction fires, and this results in an update to the propensities of reactions $k$ and $\ell$, then, if the leaves representing reactions $k$ and $\ell$ are siblings, the updates to the non-leaf nodes between $k$ and $\ell$, and the root, can be shared. This will reduce the computational cost of updating the tree. Further, if the leaves representing $k$ and $\ell$ are `cousins', then all but one of the updates to the non-leaf nodes between $k$ and $\ell$, and the root, can be shared. In order to focus on such computational improvements, at this stage, we will restrict attention to balanced trees, which means that the distance from the root to different leaves differs by at most one. We illustrate the benefits of our proposed improvements with an example.

\begin{figure}[ht]

\hrulefill
\vspace{1mm}

 \centering
 \begin{minipage}{.49\textwidth}\centering \begin{tikzpicture}[  level 1/.style={sibling distance=35.92mm},
  level 2/.style={sibling distance=20.92mm},
  level 3/.style={sibling distance=12.42mm},
  level 4/.style={sibling distance=5.42mm}], edge from parent/.style={draw,-latex}]
 \node {$\alpha = 7$}  child { node {$\alpha = 5$} child {node {$\alpha = 3$} child {node {$\alpha = 1$}} child {node {${\alpha = 2}$}} } child {node {$\vphantom{\alpha = 1}\cdot$} child {node {$\vphantom{\alpha = 1}\cdot$}} child {node {$\vphantom{\alpha = 1}\cdot$}}}} child {node {$\vphantom{\alpha = 1}\cdot$} child {node {$\vphantom{\alpha = 1}\cdot$} child {node {$\vphantom{\alpha = 1}\cdot$}} child {node {$\vphantom{\alpha = 1}\cdot$}}} child {node {$\vphantom{\alpha = 1}\cdot$} child {node {$\vphantom{\alpha = 1}\cdot$}} child {node {$\vphantom{\alpha = 1}\cdot$}}}}  ;  
 \end{tikzpicture}
\end{minipage}\hfill \begin{minipage}{.49\textwidth}\centering \begin{tikzpicture}[  level 1/.style={sibling distance=35.92mm},
  level 2/.style={sibling distance=20.92mm},
  level 3/.style={sibling distance=12.42mm},
  level 4/.style={sibling distance=5.42mm}], edge from parent/.style={draw,-latex}]
 \node {$\alpha = 7$}  child { node {$\alpha = 5$} child {node {$\alpha = 3$} child {node {$\alpha = 1$}} child {node {$\vphantom{\alpha = 1}\cdot$}} } child {node {$\vphantom{\alpha = 1}\cdot$} child {node {$\vphantom{\alpha = 1}\cdot$}} child {node {$\vphantom{\alpha = 1}\cdot$}}}} child {node {$\alpha = 6$} child {node {$\vphantom{\alpha = 1}\cdot$} child {node {$\vphantom{\alpha = 1}\cdot$}} child {node {$\vphantom{\alpha = 1}\cdot$}}} child {node {$\alpha = 4$} child {node {$\vphantom{\alpha = 1}\cdot$}} child {node {$\alpha = 2$}}}}  ;  
 \end{tikzpicture}

\end{minipage}
\caption{Two decision trees of a stochastic model comprising eight reactions are shown. If the weights of nodes $\alpha = 1$ and $\alpha = 2$ need to be recalculated, additional updates are required, and the respective values of $\alpha$ are indicated on the respective decision trees. Dots indicate the nodes that do not need to be updated. The tree on the left-hand side can be updated more efficiently than the tree on the right-hand side. Further details are contained within the main text.} \label{__label__f2d9dfa6650446a4ad861136457bf8a6}

\hrulefill
\end{figure}
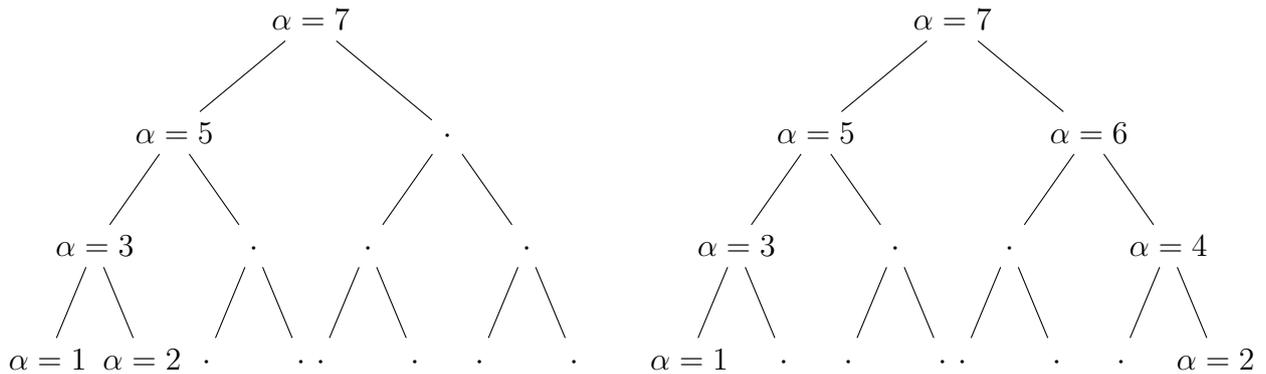

\textbf{Example.} We consider two decision trees that are shown in Figure \ref{__label__f2d9dfa6650446a4ad861136457bf8a6}. In this example, leaves with a path distance of two between them share parent-, grand-parent- and root- nodes. Leaves with a path distance of four between them share a grand-parent and the root node, whilst leaves with a path distance of six share only the root.

Suppose the $j$-th reaction fires, resulting in an update to the propensities of reactions $1$ and $2$. Thus, the weights of nodes $\alpha = 1$ and $\alpha = 2$ are recalculated. 

Recursively, all nodes on the path from $\alpha = 1$ or $\alpha = 2$ to the root need to be recalculated. If $\alpha = 1$ and $\alpha = 2$ are siblings, then there are three non-leaf nodes that need their weights recalculated: the diagram on the left shows that $\alpha = 3$, $\alpha = 5$ and $\alpha = 7$ all need to be updated. By contrast, if $\alpha = 1$ and $\alpha = 2$ are not siblings, then more internal nodes require weight recalculation: the decision tree on the right shows that $\alpha = 3$, $\alpha = 4$, $\alpha = 5$, $\alpha = 6$ and $\alpha = 7$ need their weights recalculated.  \hfill $\blacksquare$

The method for building a decision tree comprises the following two key steps: \begin{enumerate}
\item generate an \emph{interactivity graph}, with nodes representing reactions and edge weights indicating a score that captures the frequency with which pairs of reaction propensities are simultaneously updated;
\item construct a decision tree with reference to the interactivity graph, so that Algorithm \ref{__label__a13dc06ab0a6434289b4a43f9f6e09cd} can carry its task out with only limited computational resources.
\end{enumerate}

\textbf{Generating an interactivity graph.} The interactivity graph $G(V, E)$ is constructed as follows. Each reaction is represented as a node. Each pair of nodes (i.e. reactions) $(i, j)$ is connected by a weighted edge; the weight of this edge indicates the frequency of simultaneous updates to the propensities of this pair of reactions. 

The frequency of simultaneous updates to reaction propensities is not generally known \emph{a priori}, and must be estimated. A very small number of survey sample paths are generated to provide this estimate. Let $\omega_k$ indicate the total number of times reaction $k$ has fired (taken over the survey sample paths). Then, the edge weight connecting nodes $i$ and $j$ is given by \begin{equation*}
E(i, j) =  \sum_{k} \omega_k \cdot \mathbb{I}\{R_k \text{ updates } a_i\} \cdot \mathbb{I}\{R_k \text{ updates } a_j\},
\end{equation*} where the sum is taken over all reactions of the model. Alternatively, edge weights can be provided by an unweighted sum: \begin{equation*}
E(i, j) = \sum_{k} \mathbb{I}\{R_k \text{ updates } a_i\} \cdot \mathbb{I}\{R_k \text{ updates } a_j\}.
\end{equation*} The unweighted sum proceeds as if all reactions have roughly equal propensities, and has the benefit of not requiring the generation of survey simulations to calibrate the algorithm. 

\begin{figure}[hb] 
\hrulefill
\vspace{1mm}

\centering \begin{minipage}{.49\textwidth}\centering
 \begin{align*}
    \alpha =  1: \hspace{12pt} & S + I \xrightarrow{\theta_1} E + I \\
    \alpha =  2: \hspace{12pt} & E \xrightarrow{\theta_2} I \\
    \alpha =  3: \hspace{12pt} & I \xrightarrow{\theta_3} R \\
   \end{align*}

\end{minipage} \hfill \begin{minipage}{.49\textwidth}\centering\begin{tikzpicture}
 \node (c) at (3.5, 1.5) [circle,draw] {$\alpha = 1$};
 \node (a) at (7, 0)  [circle,draw] {$\alpha = 2$};
 \node (b) [circle,draw] {$\alpha = 3$};

\draw (a) -- (b) node [midway, below, sloped] (TextNode) {$  \omega_2 $};
\draw (c) -- (b) node [midway, above, sloped] (TextNode) {$ \omega_2 + \omega_3$};
\draw (a) -- (c) node [midway, above, sloped] (TextNode) {$ \omega_1 + \omega_2 $};

\end{tikzpicture}

\end{minipage}
\caption{The reactions of an S-E-I-R model are shown on the left, and a corresponding interactivity graph is shown on the right. Further details are contained within the main text.} \label{__label__9eb500d9e6a8415ab70d0ef825972a49}

\hrulefill

\end{figure}
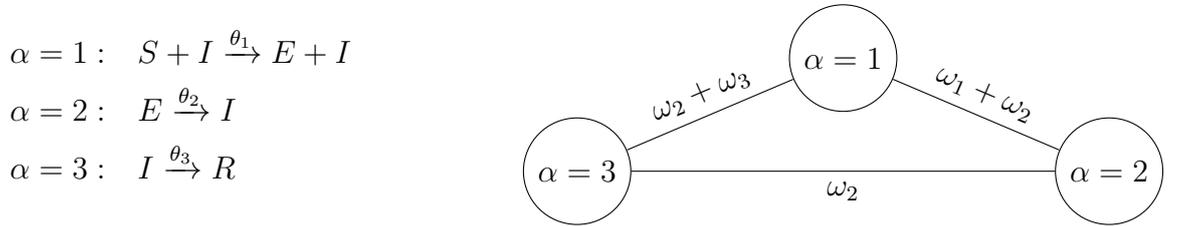

\textbf{Example.} Returning to the example of an S-E-I-R model~\citep{__ref__f498db3c1e964163a8a1231349bce15e} (see Section 3.1), on the left of Figure \ref{__label__9eb500d9e6a8415ab70d0ef825972a49} three reactions are indicated. On the right of this figure, the interactivity graph is shown: each node is associated with the indicated reaction, and the edge weights are indicated. 

\textbf{Constructing a decision tree.} Having restricted ourselves to a balanced decision tree, for a total of $M$ reactions, the number of functionally distinct decision trees is of the order of $\mathcal{O}(M! / 2^{M-1})$, which means that exhaustively searching for a decision tree is generally not feasible. Therefore, our method of constructing a binary decision tree is, by necessity, heuristic. The procedure starts by algorithmically bisecting the interactivity graph, placing each node in either the `left-' or the `right-' set, with this bisection chosen to minimise the edge cut, that is, to achieve \begin{equation*}
\min  \hphantom{x} \sum_{i \in \text{left}} \hphantom{x} \sum_{j \in \text{right}}  \hphantom{x} E(i, j), 
\end{equation*} where $E(i, j)$ is the weight of the edge between $i$ and $j$, subject to the left- and right-sets having the same size (or, if necessary, differing by one node). This means that, on the whole, where reactions $i$ and $j$ frequently have their propensities updated together, they are both in the same set. 

This partition generates the first decision of the decision tree: from the root, the left-child represents all reactions assigned to the left-set, whilst the right-child represents all reactions assigned to the right-set. The procedure is then recursively repeated inside the left-set and the right-set, until the sets can no longer be bisected. To illustrate we consider an example.

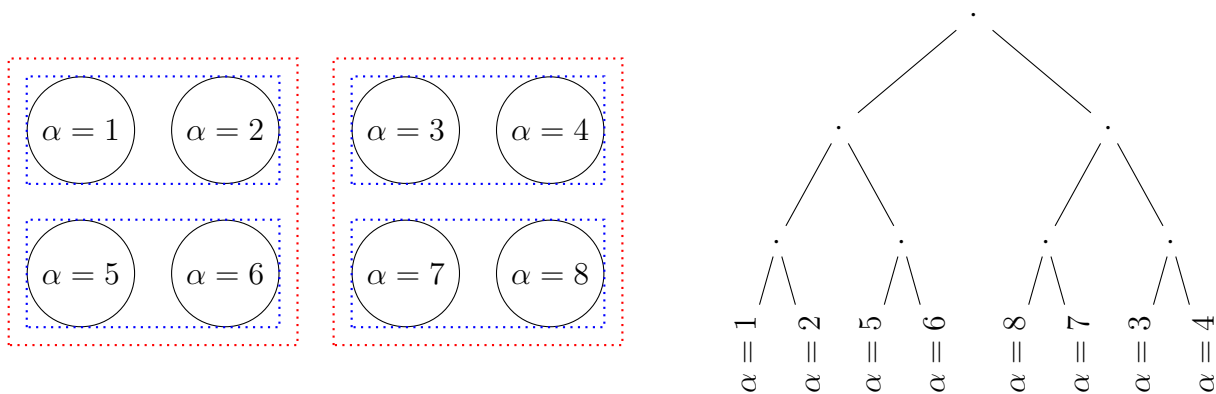
\begin{figure}[ht]

\hrulefill
\vspace{1mm}

\centering\begin{minipage}{.49\textwidth}\centering\begin{tikzpicture}[scale=.95]
 
 \node (c) at (0, 0) [circle,draw] {$\alpha = 1$};
 \node (c) at (2, 0) [circle,draw] {$\alpha = 2$};
 \node (c) at (4.5, 0) [circle,draw] {$\alpha = 3$};
 \node (c) at (6.5, 0) [circle,draw] {$\alpha = 4$};
 \node (c) at (0, -2) [circle,draw] {$\alpha = 5$};
 \node (c) at (2, -2) [circle,draw] {$\alpha = 6$};
 \node (c) at (4.5, -2) [circle,draw] {$\alpha = 7$};
 \node (c) at (6.5, -2) [circle,draw] {$\alpha = 8$};
 
 \draw[red,thick,dotted] (-1, -3)  rectangle (3, 1);
 \draw[red,thick,dotted] (3.5, -3)  rectangle (7.5, 1);
  
 \draw[blue,thick,dotted] (-0.75, -0.75)  rectangle (2.75, 0.75);
 \draw[blue,thick,dotted] (-0.75, -2.75)  rectangle (2.75, -1.25);
 
 \draw[blue,thick,dotted] (4.5-0.75, -0.75)  rectangle (4.5+2.75, 0.75);
 \draw[blue,thick,dotted] (4.5-0.75, -2.75)  rectangle (4.5+2.75, -1.25);

\end{tikzpicture}
\end{minipage}\hfill\begin{minipage}{.49\textwidth}\centering\begin{tikzpicture}[  level 1/.style={sibling distance=35.42mm},
  level 2/.style={sibling distance=16.42mm},
  level 3/.style={sibling distance=8.42mm},
  level 4/.style={sibling distance=6.42mm}], edge from parent/.style={draw,-latex}]
 \node[rotate=90]{$\cdot$}  child { node[rotate=90]{$\cdot$} child {node[rotate=90]{$\cdot$} child {node[rotate=90] {$\alpha = 1$}} child {node[rotate=90]{$\alpha = 2$}} } child {node[rotate=90]{$\cdot$} child {node[rotate=90]{$\alpha = 5$}} child {node[rotate=90]{$\alpha = 6$}}}} child {node[rotate=90]{$\cdot$} child {node[rotate=90]{$\cdot$} child {node[rotate=90]{$\alpha = 8$}} child {node[rotate=90]{$\alpha = 7$}}} child {node[rotate=90]{$\cdot$} child {node[rotate=90]{$\alpha = 3$}} child {node[rotate=90]{$\alpha = 4$}}}}  ;  
 \end{tikzpicture}

\end{minipage}

\caption{The left side indicates how an interactivity graph is repeatedly bisected. The right side shows the resultant decision generated by the algorithm. Further details are contained within the main text.} \label{__label__00bbc5b68db2454b92d1e652e72f8f4d}

\hrulefill

\end{figure}

\textbf{Example.} We consider a model comprising eight reactions. An interactivity graph is generated. Then, the eight reactions are bisected into two sets of size four. Each set of size four is bisected again. On the left of Figure \ref{__label__00bbc5b68db2454b92d1e652e72f8f4d}, the red boxes indicate the two sets after the first bisection; the blue boxes indicate the second-level bisections (and the third-level bisections are trivial). The resultant binary decision tree is shown on the right of Figure \ref{__label__00bbc5b68db2454b92d1e652e72f8f4d}. Note that the designation of left- and right-set is arbitrary and does not change the graph. \hfill $\blacksquare$

Finally, we need to specify the heuristic graph partitioning algorithm that we wish to use. The algorithm will need to, at each level, bisect the nodes whilst minimising the resultant edge cut. A wide range of algorithms would be suitable, but we will use the ubiquitous Kernighan-Lin algorithm~\citep{__ref__fac924b4dc2e408a8a3294754310fc60}; the method is detailed in Algorithm \ref{__label__d75ddc4cd05a43b2b4f1b2f6767cbdde}.

\renewcommand{\thealgorithm}{3}
 \begin{algorithm}[!b]
 \caption{Kernighan-Lin graph bisection algorithm~\citep{__ref__fac924b4dc2e408a8a3294754310fc60}.\protect\vphantom{$A_A^A$}}
 \label{__label__d75ddc4cd05a43b2b4f1b2f6767cbdde}
 
  \begin{algorithmic}[1]
  \Require interactivity graph, $G(V,E)$ \protect\vphantom{$A_A^A$}
  \State randomly partition $V$ into disjoint sets $A$ and $B$ such that $\big||A| - |B| \big|\le 1$
   \Repeat
   \State compute $D_v$ for all $v \in V$ as 
   \begin{equation*}
    D_v = \sum_{v' \in B} E(v, v') - \sum_{v' \in A} E(v, v')
   \end{equation*}

   \State let $g_v$, $a_v$, and $b_v$ be empty lists
   \For{$n = 1, \dots, \lfloor|V|/2\rfloor$}
   \State find $a \in A$ and $b \in B$, such that $g = D_a + D_b - 2\cdot E(a, b)$ is maximal
  \State remove $a$ and $b$ from further consideration in this pass
  \State           add $g$ to $g_v$, $a$ to $a_v$, and $b$ to $b_v$
  \State           update $D_v$ for $A \leftarrow A \setminus a$, and $B \leftarrow B \setminus b$
  \EndFor
   \State find $k$ which maximises $g_{max}$, the sum of $g_v[1],\dots,g_v[k]$
   \If{$g_{max} > 0$}
   \State exchange $a_v[1], a_v[2], \dots, a_v[k]$ with $b_v[1], b_v[2], \dots, b_v[k]$
  \EndIf    
 \Until $g_{max} \le 0$
 \State \Return {G(V,E)}
 
  \end{algorithmic}
 \end{algorithm}

 We are now in a position to commence numerical investigations into the new simulation method.

\section{Numerical investigations}  \label{__label__d8606a0a95494d729a9d596e0819d59d}
In this section, we present two case studies that demonstrate the efficiency of the refined simulation algorithm. The first case study is concerned with the diffusion of a single species on a patch network. The second case study models the spread of a disease between communities. 

The numerical performance of different simulation algorithms is compared by referring to the number of reactions fired per second of CPU time. We will compare the performance of a bespoke decision tree (BDT) chosen by the method described in Section \ref{__label__28abc88440d1434db601b26de0e65d7f}, with the performance of a random decision tree (RDT). The RDT will be binary and balanced. A discussion of the results follows in Section \ref{__label__47c2e02981814fe88813dc92598acc9a}. All the sample paths were generated on an Intel Core i5 CPU rated at 2.5 GHz. 

\subsection{Diffusion on a contact network} \label{__label__ad7308494b5d46f79c4299e2c0c3bbe2}
In this first case study, we consider the diffusion of a single species, $X$. The species inhabits a domain comprising $K$ patches, and individual members of the species can move from one patch to another patch only if the patches are connected. 

Clearly, the relative computational performance of competing simulation algorithms will depend on the `contact network' that defines which patches are connected to each other. Suppose each patch of the model is associated with a node in the contact network. If it is possible to move from one patch to another in the model, there is to be an edge between the associated nodes of the contact network. We will generate the interaction network by following two widely-used frameworks: the Erd\H{o}s-R{\'e}nyi~\citep{__ref__f10017de4dbe4fce9d3c5e89d050044d} and Barab{\'a}si-Albert~\citep{__ref__3cd123e61efb41538d737672bde681ea} models. Each framework is now introduced. 

\textbf{The Erd\H{o}s-R{\'e}nyi model.} An Erd\H{o}s-R{\'e}nyi (ER) model on $K$ nodes \citep{__ref__f10017de4dbe4fce9d3c5e89d050044d} is constructed by connecting nodes randomly. Each out of the possible edges is included with constant probability $p$; different edges are independently included.

\textbf{The Barab{\'a}si-Albert model}. The Barab{\'a}si-Albert (BA) model is a widely-used example of a scale-free interaction network that generates a random graph on $K$ nodes, and is in widespread use as a modelling tool. For example, the BA model has been used to describe hyperlinks between web-pages of the internet~\citep{__ref__3cd123e61efb41538d737672bde681ea}. The BA model uses preferential attachment, which means that, as new nodes are added to a graph, they are more likely to be connected to existing nodes with more edges radiating from them. \citet{__ref__3cd123e61efb41538d737672bde681ea} describe their model as follows:

\begin{quotation}
``Starting with a small number ($m_0$) vertices [nodes], at every time step we add a new vertex with $m (\le m_0)$ edges that link the new vertex to $m$ different vertices already present in the system. To incorporate preferential attachment, we assume that the probability $\Pi$ that a new vertex will be connected to a vertex $i$ depends on the connectivity $k_i$ of that vertex, so that $\Pi(k_i) = k_i / \sum_j k_j$.''
\end{quotation}

In order to model the movement of a species, $X$, over $K$ patches, we fix $K = 50$. To robustly compare competing simulation algorithms, we must first generate different contact networks, and then compare the performance of the competing algorithms on each of the contact networks.

We start with the ER model. First, we set $p \in \{1/10, 3/10, 5/10, 7/10, 9/10\}$: larger values of $p$ result in a higher proportion of patches being in contact with other patches. Then, for each value of $p$, five different stochastic models are generated, each having its own contact network. Thus, there are $25$ models in total. For each of the $25$ models, we proceed to: \begin{itemize}
\item construct a BDT tree according to Section \ref{__label__28abc88440d1434db601b26de0e65d7f};
\item construct $100$ RDTs by randomly assigning reactions to leaves of the tree.
\end{itemize} We then compare the performance of the simulation algorithm with a carefully-chosen BDT against a reference distribution of RDTs. 

\begin{figure}[b]
 \centering
 \includegraphics[width=\linewidth]{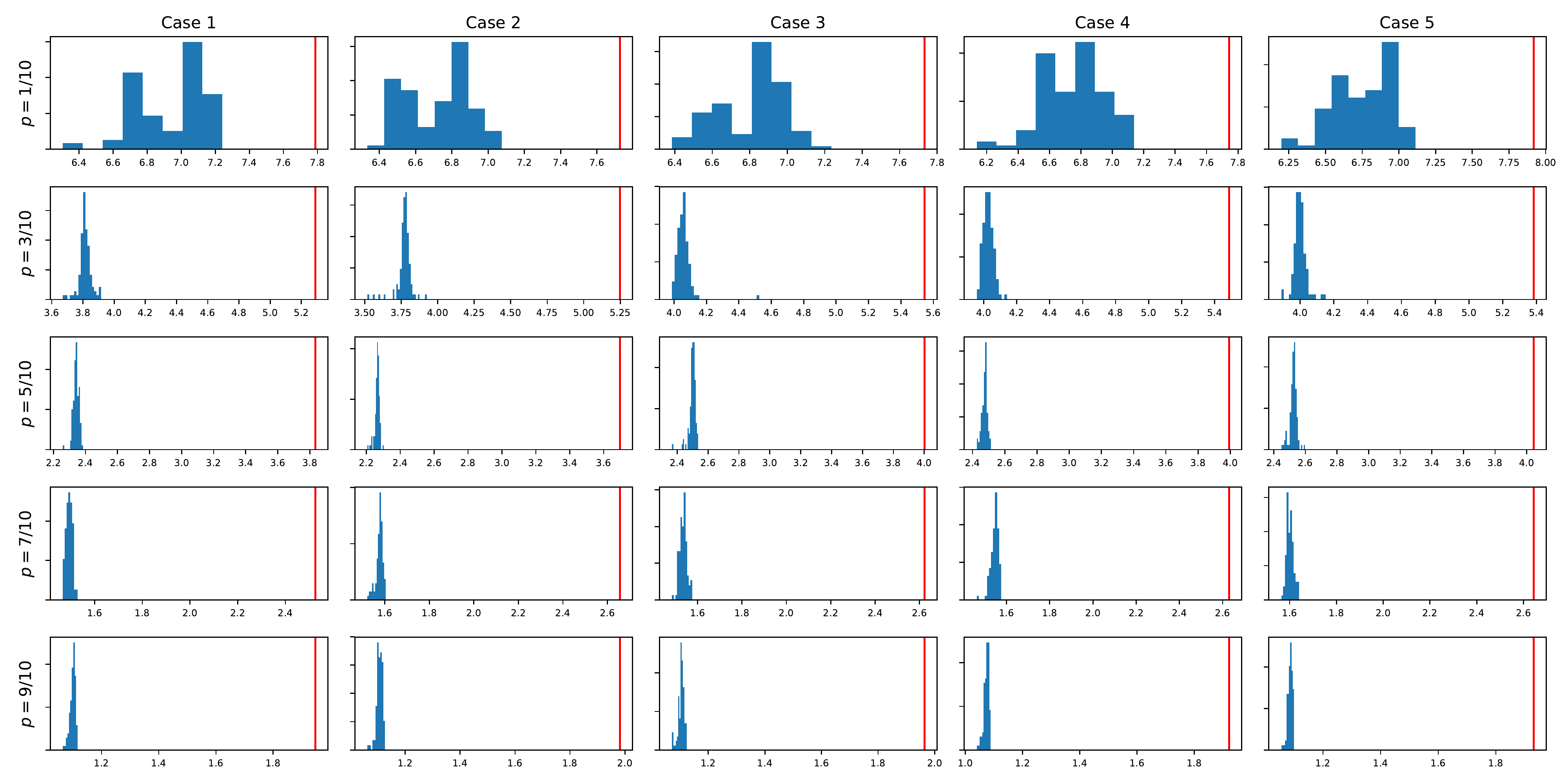}
 \caption{Twenty-five contact-networks are generated, and the movement of particles, as described in Section \ref{__label__ad7308494b5d46f79c4299e2c0c3bbe2}, on these networks is simulated. The computational performance of a BDT is compared with the performance of RDTs.  The histograms represent the average number of reactions fired per CPU second for each of the $100$ different RDTs, whilst the red lines indicate the reaction rate of the BDT. The contact networks are ER graphs generated with the indicated parameter value of $p$.} \label{__label__6bf67b046efe483db77ec025ef3ac7b2}
\end{figure}

\begin{figure}[htb]
 \centering

 \includegraphics[width=\linewidth]{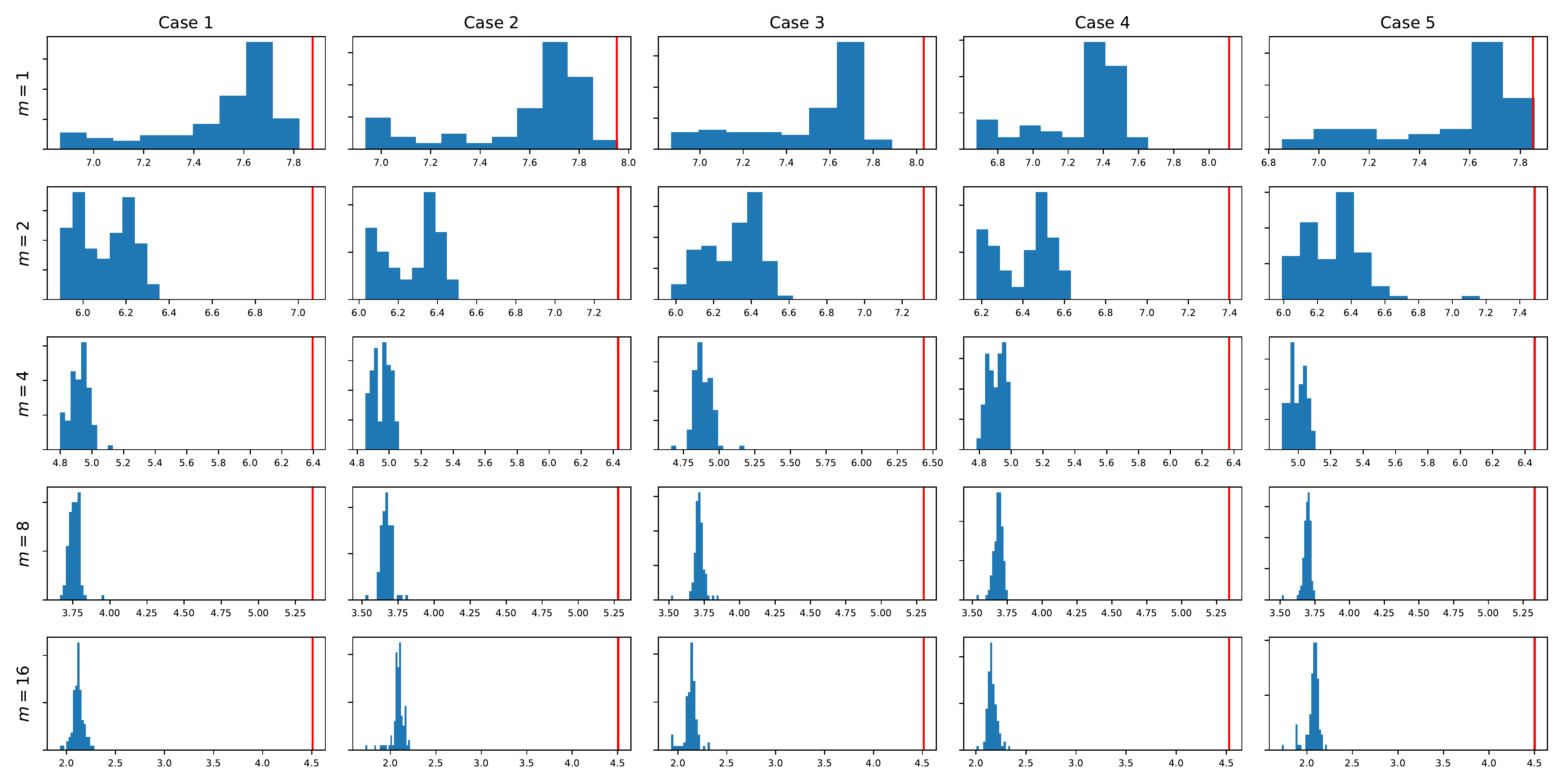}
 \caption{Twenty-five contact-networks are generated, and the movement of particles on these networks, as described in Section \ref{__label__ad7308494b5d46f79c4299e2c0c3bbe2}, is simulated. The computational performance of a BDT is compared with the performance of RDTs.  The histograms represent the average number of reactions fired per CPU second for each of the $100$ different RDTs, whilst the red lines indicate the reaction rate of the BDT. The contact networks are BA graphs generated with the indicated parameter value of $m$.} \label{__label__4428e69251b94c71b113b71f50a260ad}
\end{figure}

In Table \ref{__label__8e376c66750e488f9015ed629098996d} and Figure \ref{__label__6bf67b046efe483db77ec025ef3ac7b2}, we compare the computational performance of a BDT generated as described in Section \ref{__label__28abc88440d1434db601b26de0e65d7f} against the performance of random RDTs. For each of the aforementioned choices of $p$, and for each of the five randomly-generated ER contact networks, we generate a BDT. We then generate $100$ RDTs. The decision trees are then used to simulate sample paths; we determine the number of reactions simulated by each algorithm per CPU-second. The summary provided by Table \ref{__label__8e376c66750e488f9015ed629098996d} shows that a BDT  typically outperforms an RDT, with the BDT being able to simulate between 17\% and 330\% more reactions per CPU-second (with the speed-up factor depending on the value of $p$).  Figure \ref{__label__4428e69251b94c71b113b71f50a260ad} provides more detail: for each different contact network, the CPU performance of each of the 100 RDTs is represented as a histogram, and the performance of the BDT is superimposed. In particular, Figure \ref{__label__4428e69251b94c71b113b71f50a260ad} demonstrates that the BDT consistently outperforms RDTs. The improvements introduced with a BDT are more apparent for larger values of $p$: we return to this point in the discussion (Section \ref{__label__47c2e02981814fe88813dc92598acc9a}).

\begin{table}[t]
\centering\hrulefill\vspace{3mm}

\begin{tabular}{|l|ccccc|} \hline\hline
\rule{0pt}{12pt}\textbf{Configuration} & $p = 1/10$ & $p = 3/10$ & $p = 5/10$ & $p = 7/10$ & $p = 9/10$\\ \hline
\rule{0pt}{12pt}\textbf{Bespoke tree ($s^{-1}$)} & $7.20 \cdot 10^6$ & $5.28 \cdot 10^6$ &$4.42 \cdot 10^6$ &$3.78 \cdot 10^6$ &$3.23 \cdot 10^6$ \\ 
\rule{0pt}{12pt}\textbf{Random tree ($s^{-1}$)} & $6.13 \cdot 10^6$ & $3.71 \cdot 10^6$ &$1.91 \cdot 10^6$ &$1.03 \cdot 10^6$ &$0.75 \cdot 10^6$ \\ \hline
\rule{0pt}{12pt}\textbf{Speed-up (\%)} & 17 & 42 & 132 & 266 & 330\\\hline\hline
\end{tabular}

\caption{The computational performance of a BDT is compared with the performance of RDTs. This table shows the average number of reactions fired per CPU-second when movement on a contact-network based on an ER model is simulated. The averages are calculated for each choice of the parameter $p$ of the ER model. The BDT is significantly more efficient, especially for densely connected patch networks.} \label{__label__8e376c66750e488f9015ed629098996d}
\hrulefill
\end{table}

We now proceed to consider the BA model. As before, a range of interaction graphs is generated by sweeping through different choices of $m$: we take $m \in \{1, 2, 4, 8, 16\}$, and $m_0 = m$. For each choice of $m$, we generate five random interaction graphs as a basis for comparing the BDT with a range of RDTs. In Table \ref{__label__2a8ab29afa4841d9b16ddd1248bb2c68}, the computational performance of the different decision trees is compared. Table \ref{__label__2a8ab29afa4841d9b16ddd1248bb2c68} shows that, based on the test data, a BDT is, on average, between 7\% and 114\% more efficient than an RDT. The efficiencies are more apparent when $m$ is large: indeed, this corresponds with a greater number of patches being in contact with each other. 

Further details are provided in Figure \ref{__label__4428e69251b94c71b113b71f50a260ad}. We point out that Figure \ref{__label__4428e69251b94c71b113b71f50a260ad} illustrates that when very few edges are present in the contact network the performance of the BDT is not dissimilar to the typical performance of an RDT. Where there are few edges in the contact network the interactivity graph is sparse and there are only limited opportunities to simultaneously update the weights of the nodes of the decision tree. As such, only limited computational savings over an RDT are possible.

\begin{table}[t]
\hrulefill
\centering
\vspace{8mm}

\begin{tabular}{|l|ccccc|} \hline\hline
\rule{0pt}{12pt}\textbf{Configuration} & $m = 1$ & $m = 2$ & $m = 4$ & $m = 8$ & $m = 16$\\
\rule{0pt}{12pt}\textbf{Edges present (\%)} & 4 & 8 & 15 & 27 & 44 \\ \hline
\rule{0pt}{12pt}\textbf{Bespoke tree ($s^{-1}$)} & $7.96 \cdot 10^6$ & $7.31 \cdot 10^6$ &$6.42 \cdot 10^6$ &$5.32 \cdot 10^6$ &$4.51 \cdot 10^6$ \\ 
\rule{0pt}{12pt}\textbf{Random tree ($s^{-1}$)} & $7.47 \cdot 10^6$ & $6.27 \cdot 10^6$ &$4.93 \cdot 10^6$ &$3.70 \cdot 10^6$ &$2.11 \cdot 10^6$ \\ \hline
\rule{0pt}{12pt}\textbf{Speed-up (\%)} & 7 & 17 & 30 & 44 & 114\\\hline\hline
\end{tabular}
\caption{The computational performance of a BDT is compared with the performance of RDTs. This table shows the average number of reactions fired per CPU-second when movement on a contact-network based on an BA model is simulated. The averages are calculated for each choice of the parameter $m$ of the BA model, and the proportion of edges present in the contact network is also shown. The BDT is generally more efficient, especially for densely connected patch networks.}\label{__label__2a8ab29afa4841d9b16ddd1248bb2c68}

\hrulefill

\end{table}

\subsection{Disease spread on a contact network} \label{__label__3d517af24a2c4890a5fe484f9f3893d9}
In this second case study, a spatially inhomogeneous S-I-S compartment model is investigated~\citep{__ref__f498db3c1e964163a8a1231349bce15e}. Individuals within this model are either susceptible (state `S') or infected (state `I') with a disease, and the patch that contains the individual is recorded. An infection of a susceptible individual in patch $k$ can occur due to either a local or a long-range infection. A local infection is represented as the reaction $$S_k + I_k \xrightarrow{\theta_1} 2 \cdot I_k.$$ A contact network specifies how long-range infections take place: where patches $k$ and $\ell$ are in contact, a long-range infection is represented as $$S_k + I_\ell \xrightarrow{\theta_2} I_k + I_\ell\,\,\,(\text{for } k \neq \ell).$$ The recovery of an infected individual can be represented as $$I_k \xrightarrow{\theta_3} S_k.$$ We take $(\theta_1, \theta_2, \theta_3)$ as $(1, 10^{-2}, 10^2)$, and, inside each of the $K = 50$ patches, initially there are $10$ infected and $90$ susceptible individuals. Contact networks, that indicate which long-range infections are possible, are generated with the ER method with the parameter $p$ set as  $p \in \{1/10, 3/10, 5/10, 7/10, 9/10\}$. Then for each value of $p$ five different stochastic models are generated, each having its own contact network.

\begin{figure}[ht]
 \centering
  \includegraphics[width=\linewidth]{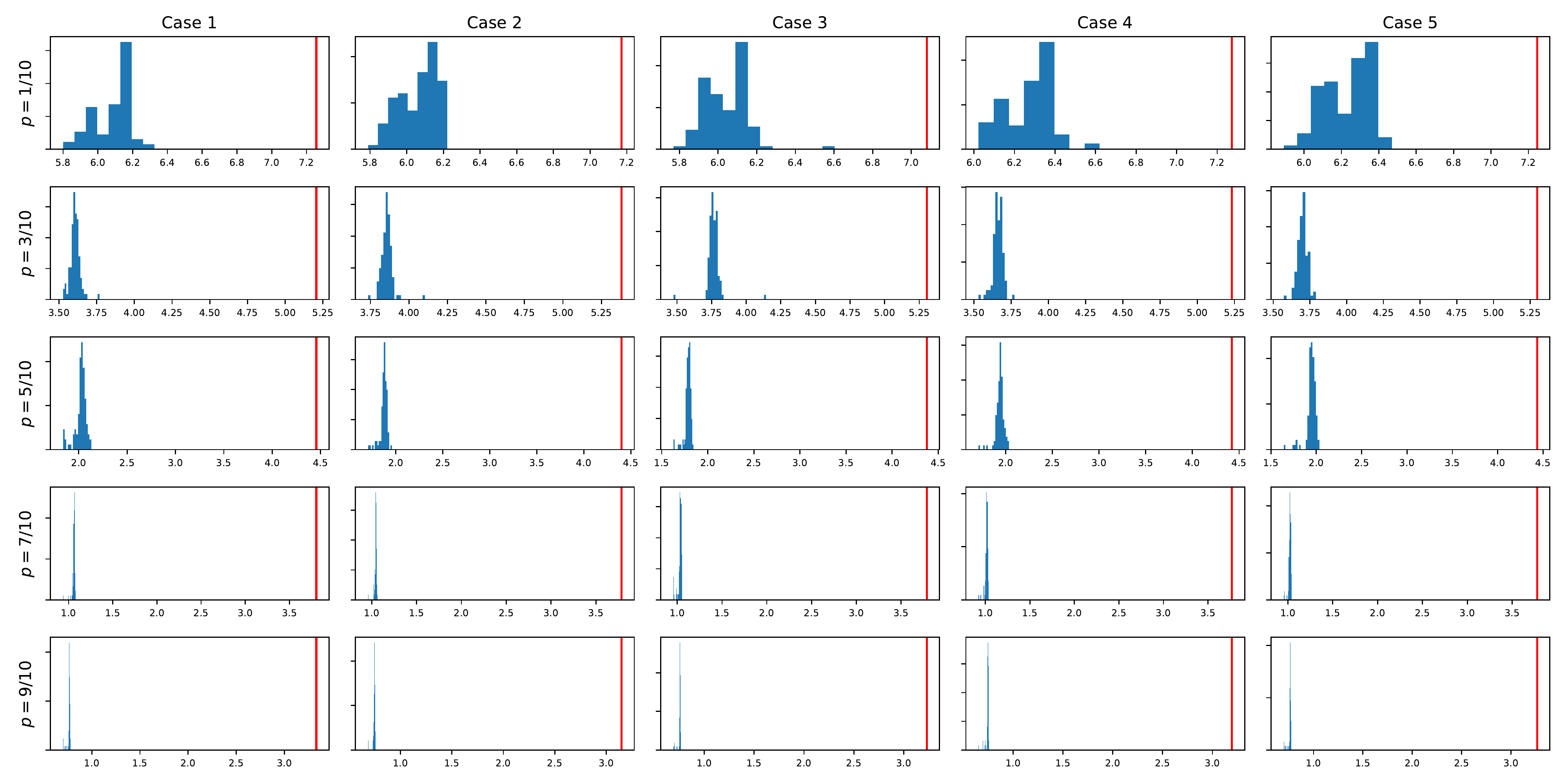}
 \caption{Twenty-five contact-networks are generated, and the spread of a disease on these networks, as described in Section \ref{__label__3d517af24a2c4890a5fe484f9f3893d9}, is simulated. The computational performance of a BDT is compared with the performance of RDTs.  The histograms represent the average number of reactions fired per CPU second for each of the $100$ different RDTs, whilst the red lines indicate the reaction rate of the BDT. The contact networks are BA graphs generated with the indicated parameter value of $p$.} \label{__label__bf1a93aead704eeeb64baf01652a048c}
\end{figure}

As before, we compare the performance of a BDT generated as described in Section \ref{__label__28abc88440d1434db601b26de0e65d7f} against a reference distribution of RDTs. In Table \ref{__label__e0a8b52d63544d94af71508330b95c95}, the computational performance of the different decision trees is compared. Table \ref{__label__e0a8b52d63544d94af71508330b95c95} shows that, based on the test data, a BDT is, on average, between 15\% and  78\% more efficient than an RDT. Further information is contained in Figure \ref{__label__bf1a93aead704eeeb64baf01652a048c}. As with the case study presented in Section \ref{__label__ad7308494b5d46f79c4299e2c0c3bbe2}, when $p$ is small, very few edges are present in the model, and in these cases, Figure \ref{__label__bf1a93aead704eeeb64baf01652a048c} confirms that the performance of the BDT is not dissimilar to the typical performance of an RDT.

\begin{table}[ht]
\centering\hrulefill\vspace{3mm}

\begin{tabular}{|l|ccccc|} \hline\hline
\rule{0pt}{12pt}\textbf{Configuration} & $p = 1/10$ & $p = 3/10$ & $p = 5/10$ & $p = 7/10$ & $p = 9/10$\\ \hline
\rule{0pt}{12pt}\textbf{Bespoke tree ($s^{-1}$)} & $7.78 \cdot 10^6$ & $5.38 \cdot 10^6$ &$3.90 \cdot 10^6$ &$2.61 \cdot 10^6$ &$1.95 \cdot 10^6$ \\ 
\rule{0pt}{12pt}\textbf{Random tree ($s^{-1}$)} & $6.78 \cdot 10^6$ & $3.92 \cdot 10^6$ &$2.41 \cdot 10^6$ &$1.55 \cdot 10^6$ &$1.09 \cdot 10^6$ \\ \hline
\rule{0pt}{12pt}\textbf{Speed-up (\%)} & 15 & 37 &  62 &  69 &  78\\\hline\hline
\end{tabular}

\caption{The computational performance of a BDT is compared with the performance of RDTs. This table shows the average number of reactions fired per CPU-second where the long-range interactions of an S-I-S model are determined with an ER-modelled contact network. The averages are calculated for each choice of the parameter $p$ of the ER model. The BDT is generally more efficient.}\label{__label__e0a8b52d63544d94af71508330b95c95}
\hrulefill

\end{table}

\section{Discussion}  \label{__label__47c2e02981814fe88813dc92598acc9a}
Agent-based models of complicated real-world phenomena would serve no useful purpose if it were impossible to characterise the resultant models' dynamics. As such, the design and development of efficient stochastic simulation algorithms play a pivotal role in underpinning the work of the modelling community. In this work, we present a technique for designing simulation algorithms that exploit the specific dynamics of the stochastic, discrete-state model of interest. We then proceeded to evaluate the computational performance with a number of case studies. 

In this final section, we discuss the conclusions under three headings: firstly, we explain how a highly efficient simulation algorithm facilitates the use of more complicated and realistic discrete-state models as a modelling tool. Secondly, we explain how these conclusions will inevitably lead to the development of parallelised simulation algorithms for event-based models. Thirdly, we present our view of the direction future research is likely to take. We then conclude with an outlook. 

\textbf{Modelling implications.} By using a carefully-chosen binary decision tree to fire reactions in a simulation algorithm, we allow more sample paths to be generated for a given level of computational resources. An immediate benefit is the ability for a researcher to conduct parameter sweeps of a chosen model in more detail, thereby being in a stronger position to characterise the model dynamics. A more subtle benefit lies in the ability to describe more complicated models, since our BDTs make it feasible to include many, long-range interactions as part of the model, even if these interactions only occur infrequently.

\begin{figure}
 \centering

 \includegraphics[width=.75\linewidth]{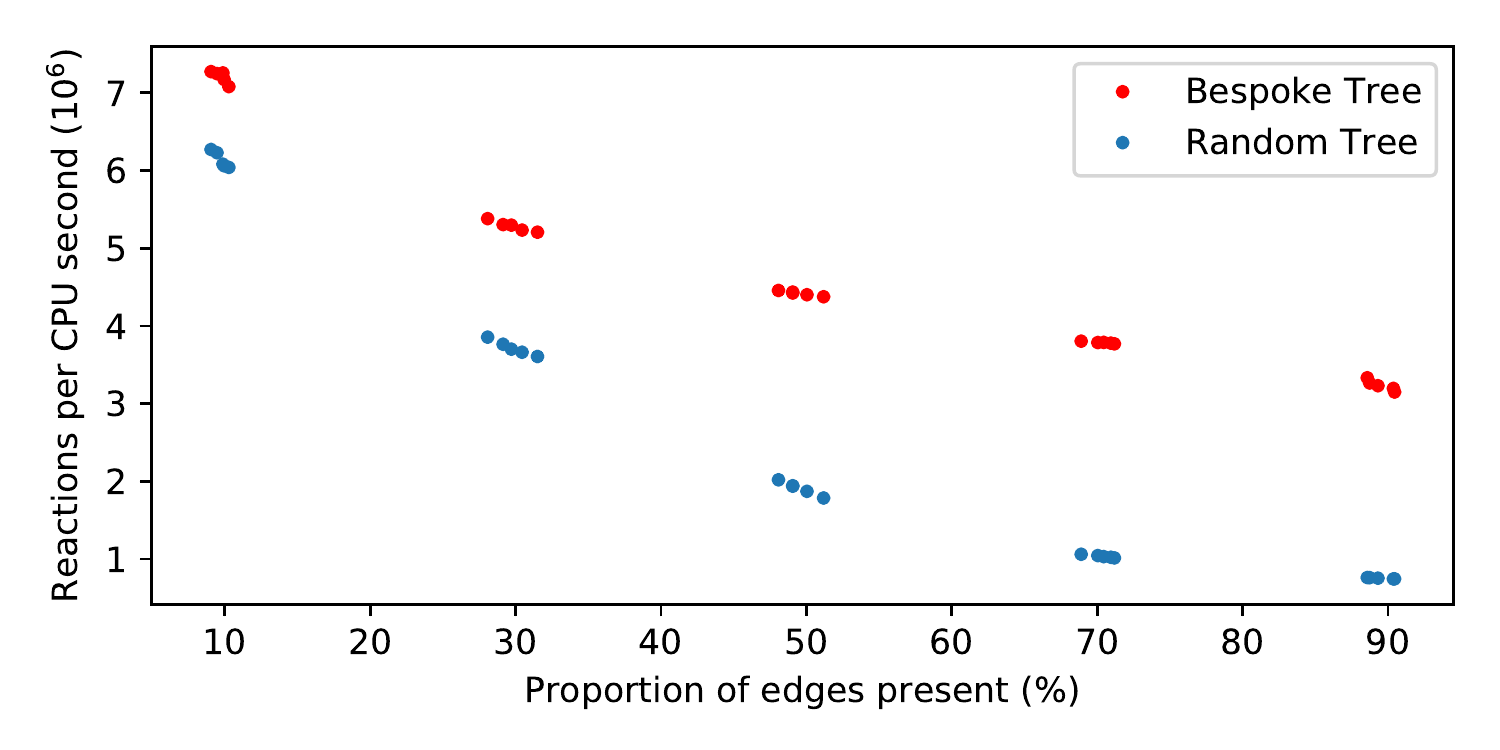}
 \caption{The average number of reactions per CPU-second is shown for BDT and RDT-based simulations of movement on a contact network. As per Section \ref{__label__ad7308494b5d46f79c4299e2c0c3bbe2}, 25 different contact networks are generated (five contacts networks for each choice of $p$). On the $x$-axis we show the proportion of edges present in each contact network, and the $y$-axis indicates the corresponding numerical performance. Whilst an increase in the number of edges slows down the simulation algorithm, the effect is more pronounced for the RDT than the BDT method.} \label{__label__f6170bae97e7487b845524172d2c045c}
\end{figure}

Section \ref{__label__d8606a0a95494d729a9d596e0819d59d} demonstrates that, for a fixed number of species and patches, this algorithm is particularly suited for use with models with a relatively high number of reactions. In particular, in Section \ref{__label__ad7308494b5d46f79c4299e2c0c3bbe2} we studied the movement of individuals on Erd\H{o}s-R{\'e}nyi contact graphs. Figure \ref{__label__f6170bae97e7487b845524172d2c045c} shows that, as the number of patch-to-patch contacts increases, the number of reactions per CPU-second of both a BDT- and an RDT-based simulation algorithm decreases. However, the performance of a bespoke algorithm is significantly less affected.

\textbf{Parallel simulation algorithms.} Whilst a parallelised implementation of a simulation algorithm will not reduce the CPU-seconds required for each sample path, by using more resources simultaneously, model statistics can be available sooner to an end-user. For simple discrete-state models, sample paths can be simultaneously generated through a trivial parallelisation: for each CPU core, a new thread, endowed with a different random number seed, is spawned and the simulation completed. Where a model comprises a very large number of interacting species, it might be necessary for the system to be partitioned, with different components of the system assigned to different CPU cores for simulation. Where a cross-partition interaction is to take place, the different CPU threads will need to be synchronised. In our view, the development of an interactivity graph represents an important advance in transforming stochastic simulation into a parallelised endeavour.

\textbf{Future work.} Additional work that builds on our bespoke simulation strategy is likely to take on three forms. Firstly, by relaxing the constraints we have imposed on the decision tree, namely, that it is both balanced and binary, further computational improvements might be possible. Secondly, inline updating of the decision tree, that is, a continuous refinement of the decision tree structure as sample paths are generated might also be possible. Thirdly, alternative approaches for generating decision trees, such as the use of a genetic algorithm, might also yield promising results.

\subsection{Outlook}
In this work, we demonstrated that a binary decision tree provides an effective means of sampling reactions of a discrete-state stochastic model. Careful choice of the decision tree provides significant computational improvements over a random selection of a decision tree. Whilst the benefits of the new method are most noticeable in a stochastic model comprising a relatively high number of reactions, the method is suitable for use with a general, discrete-state model.

\bibliographystyle{v2}
\bibliography{r1_clean.bib}

\begin{thebibliography}{19}
\providecommand{\natexlab}[1]{#1}

\bibitem[Allen(2010)]{__ref__33611e8f2913412d9e8fe5864f57d3c7}
Allen, L.~J.
\newblock \emph{{An Introduction to Stochastic Processes with Applications to
  Biology}}.
\newblock CRC Press, 2010.

\bibitem[Wilkinson(2006)]{__ref__1a3541a1fee44aae857a8870b87649ef}
Wilkinson, D.~J.
\newblock \emph{Stochastic Modelling for Systems Biology}.
\newblock CRC Press, 2006.

\bibitem[Van~Kampen(1992)]{__ref__221389c689d04b34b5c22896abc672e7}
Van~Kampen, N.~G.
\newblock \emph{{Stochastic Processes in Physics and Chemistry}}, volume~1.
\newblock Elsevier, 1992.

\bibitem[Andersson and Britton(2012)]{__ref__6b94d95a4d484ce0b244336a44f6ecea}
Andersson, H. and Britton, T.
\newblock \emph{Stochastic Epidemic Models and their Statistical Analysis}.
\newblock Springer, 2012.

\bibitem[Diekmann and
  Heesterbeek(2000)]{__ref__f498db3c1e964163a8a1231349bce15e}
Diekmann, O. and Heesterbeek, J. A.~P.
\newblock \emph{Mathematical Epidemiology of Infectious Diseases: Model
  Building, Analysis and Interpretation}, volume~5.
\newblock John Wiley \& Sons, 2000.

\bibitem[Gillespie et~al.(2013)]{__ref__115d94a32a21424e98958e07bcc499e4}
Gillespie, D.~T., Hellander, A., and Petzold, L.~R.
\newblock Perspective: Stochastic algorithms for chemical kinetics.
\newblock \emph{Journal of Chemical Physics}, \textbf{138}(17), 2013.

\bibitem[Gillespie(1976)]{__ref__95f69d794ca6465396aa1b23c5e7ad6b}
Gillespie, D.~T.
\newblock A general method for numerically simulating the stochastic time
  evolution of coupled chemical reactions.
\newblock \emph{{Journal of Computational Physics}}, \textbf{22}(4):403--434,
  1976.

\bibitem[Gillespie(1977)]{__ref__35755be6131a4c9e83117b17737c061c}
Gillespie, D.~T.
\newblock {Exact stochastic simulation of coupled chemical reactions}.
\newblock \emph{{Journal of Physical Chemistry}}, \textbf{81}(25):2340--2361,
  1977.

\bibitem[Hattne et~al.(2005)]{__ref__db66f64ab877460bb7754f2fcb5a3f3c}
Hattne, J., Fange, D., and Elf, J.
\newblock Stochastic reaction-diffusion simulation with MesoRD.
\newblock \emph{Bioinformatics}, \textbf{21}(12):2923--2924, 2005.

\bibitem[Drawert et~al.(2012)]{__ref__2be02b1393a8492faf235127e91447fd}
Drawert, B., Engblom, S., and Hellander, A.
\newblock URDME: a modular framework for stochastic simulation of
  reaction-transport processes in complex geometries.
\newblock \emph{BMC Systems Biology}, \textbf{6}(1):76, 2012.

\bibitem[Li and Petzold(2006)]{__ref__c70bde5c8fa14266b1df999c5ac5bd04}
Li, H. and Petzold, L.
\newblock Logarithmic direct method for discrete stochastic simulation of
  chemically reacting systems.
\newblock \emph{UCSB, Technical Report}, 2006.

\bibitem[Meinecke(2016)]{__ref__0006fcbb9b854a86a6496bf60d524597}
Meinecke, L.
\newblock \emph{Stochastic Simulation of Multiscale Reaction-Diffusion Models
  via First Exit Times}.
\newblock Ph.D. thesis, Uppsala University, 2016.

\bibitem[Cao et~al.(2004)]{__ref__a8995c057f7c4f3cb23c698d718f24ac}
Cao, Y., Li, H., and Petzold, L.
\newblock Efficient formulation of the stochastic simulation algorithm for
  chemically reacting systems.
\newblock \emph{Journal of Chemical Physics}, \textbf{121}(9):4059--4067, 2004.

\bibitem[McCollum et~al.(2006)]{__ref__e9a0ea25095f41ef97d1cdf874d0eaeb}
McCollum, J.~M., Peterson, G.~D., Cox, C.~D., Simpson, M.~L., and Samatova,
  N.~F.
\newblock The sorting direct method for stochastic simulation of biochemical
  systems with varying reaction execution behavior.
\newblock \emph{Computational Biology and Chemistry}, \textbf{30}(1):39--49,
  2006.

\bibitem[Yates and Klingbeil(2013)]{__ref__66d52925a8c64bbc83068374f2ed9600}
Yates, C.~A. and Klingbeil, G.
\newblock Recycling random numbers in the stochastic simulation algorithm.
\newblock \emph{Journal of Chemical Physics}, \textbf{138}(9):094103, 2013.

\bibitem[Welsh(1988)]{__ref__2118077d287a48c89870a834c2690e86}
Welsh, D.
\newblock \emph{Codes and Cryptography}.
\newblock Oxford University Press, 1988.

\bibitem[Kernighan and Lin(1970)]{__ref__fac924b4dc2e408a8a3294754310fc60}
Kernighan, B.~W. and Lin, S.
\newblock An efficient heuristic procedure for partitioning graphs.
\newblock \emph{The Bell System Technical Journal}, \textbf{49}(2):291--307,
  1970.

\bibitem[Erd\H{o}s and
  R{\'e}nyi(1960)]{__ref__f10017de4dbe4fce9d3c5e89d050044d}
Erd\H{o}s, P. and R{\'e}nyi, A.
\newblock On the evolution of random graphs.
\newblock \emph{Publications of the Mathematical Institute of the Hungarian
  Academy of Sciences}, \textbf{5}(1):17--60, 1960.

\bibitem[Barab{\'a}si and
  Albert(1999)]{__ref__3cd123e61efb41538d737672bde681ea}
Barab{\'a}si, A.-L. and Albert, R.
\newblock Emergence of scaling in random networks.
\newblock \emph{Science}, \textbf{286}(5439):509--512, 1999.

\end{thebibliography}

\newpage

\end{document}